\documentclass[10pt]{article}
\usepackage[margin=1in]{geometry}           
\usepackage{amsmath,amssymb,amsfonts,amsthm}

\usepackage{graphicx}                       
\usepackage{float}                          
\usepackage{subfigure}                     
\usepackage{booktabs}                       
\usepackage{multirow}                       
\usepackage{xcolor}                         
\usepackage{url}                            
\usepackage{setspace}                       
\usepackage{makecell}

\usepackage[utf8]{inputenc}
\usepackage[T1]{fontenc}

\usepackage{authblk} 

\usepackage[numbers,sort&compress]{natbib} 

\usepackage{subcaption}

\newtheorem{theorem}{Theorem}
\newtheorem{proposition}{Proposition}
\newtheorem{definition}{Definition}

\newtheorem{lemma}{Lemma}

\renewcommand{\eqref}[1]{(\ref{#1})}

\newcommand{\revised}[1]{\textcolor{black}{#1}}

\definecolor{resultbrown}{RGB}{150,75,45}
\DeclareRobustCommand{\resultedit}[1]{\textcolor{black}{#1}}

\newcommand{\acro}{ECO}

\begin{document}

\title{\textbf{ECO: Energy-Constrained Operator Learning for Chaotic Dynamics with Boundedness Guarantees}}

\author{Sunbochen Tang\textsuperscript{*}}
\author{Andrea Goertzen\textsuperscript{*}}
\author{Navid Azizan}
\affil{Massachusetts Institute of Technology}

\date{} 



\maketitle
\begingroup
  \renewcommand\thefootnote{*}
  \footnotetext{Equal Contribution.}
\endgroup

\begin{abstract}
    Chaotic systems, including weather systems and fluid turbulence, are difficult to predict due to their sensitivity to initial conditions. Many of them are dissipative and ergodic, motivating data-driven models that learn long-term invariant statistics. Despite recent progress, these models are prone to generating unbounded predictions, preventing meaningful statistical evaluation. We introduce the Energy-Constrained Operator (ECO), which jointly learns system dynamics and an energy function while enforcing boundedness in predictions. Drawing on control theory, we derive algebraic dissipativity conditions and enforce them through an efficient closed-form quadratic projection layer, yielding provable trajectory boundedness. To the best of our knowledge, this is the first work establishing such formal guarantees for data-driven chaotic dynamics models. The learned invariant level set also provides an outer estimate of the strange attractor, which is computationally intractable to characterize. Numerical experiments on the Kuramoto-Sivashinsky and Navier-Stokes equations show that ECO generates stable long-horizon forecasts that capture invariant statistics.
\end{abstract}

\section{Introduction}
\textbf{Chaotic dynamical systems.} Chaos arises in a wide range of physical systems, including weather models \citep{lorenz1963deterministic} and fluid dynamics \citep{kuramoto1978diffusion, ashinsky1988nonlinear}. A hallmark of chaotic systems is their sensitivity to initial conditions. That is, small perturbations can cause exponential divergence between trajectories, making precise long-term prediction intractable. Despite this unpredictability, many chaotic physical systems are dissipative, meaning their trajectories converge to a lower-dimensional invariant set, often referred to as the strange attractor \citep{stuart1998dynamical}. Once in this attractor, the system exhibits ergodicity, where the trajectory will eventually visit every state on the attractor \citep{guckenheimer2013nonlinear}. The ergodic behavior of dissipative chaotic systems, coupled with the intractability in predicting exact pointwise trajectories, makes capturing the statistical behavior of a system on the attractor a natural goal.

Data-driven methods for modeling chaotic systems often adopt an autoregressive framework, where a model predicts the next time step based on the current state to generate long-term trajectories and capture statistical properties. Recent approaches span across different levels of model complexity, including regression models \citep{majda2001mathematical, majda2012physics,brenner2023stability}, deep learning models \citep{li2020fourier, tang2024learning, raissi2019physics, brunton2022data, lu2021learning, kochkov2021machine, page2024recurrent}, and recurrent sequential models \citep{mikhaeil2022difficulty, vlachas2018data, sangiorgio2020robustness}. Although these models have made great progress, the autoregressive nature that they share makes them vulnerable to accumulated errors, causing trajectories to drift away from the training distribution. In chaotic settings, such drift can result in unbounded or nonphysical predictions, ultimately corrupting statistical estimates derived from the generated trajectories. 
Structured models, such as multi-level quadratic regression, have been theoretically shown to exhibit pathological instability in their statistical behavior \citep{majda2012fundamental}. In the context of recurrent neural networks (RNNs), \citet{mikhaeil2022difficulty} demonstrated that training such models on chaotic systems leads to gradient divergence, highlighting a fundamental limitation. While the theoretical understanding of machine learning models remains limited, empirical evidence shows that these models often produce divergent trajectories, ultimately resulting in unreliable statistical predictions. This issue has also been observed in advanced time-series modeling techniques, including Reservoir Computing and Fourier Neural Operators \citep{lu2018attractor, pathak2017using, li2022learning}. As such, a central challenge across both physics-informed and purely data-driven approaches is the difficulty of maintaining stable long-term forecasts.

\revised{
\textbf{Neural Operators.} Many physical systems of interest are governed by complex spatio-temporal behavior represented as partial differential equations (PDEs). Neural operators have emerged as surrogate representations well-suited for PDE learning due to their function-to-function mapping paradigm. That is, rather than mapping between vectors of prescribed length, neural operators map between function spaces, making them particularly useful in PDE learning, where entire spatial domains evolve in time. In practice, evaluations must be finitely discretized, but neural operators are discretization agnostic, 
where evaluation can be performed on any discretization, regardless of the training discretization. Several neural operator architectures have demonstrated success as surrogate models for complex spatio-temporal processes. The DeepONet~\cite{lu2021learning} trains two networks; one processes the input function on a fixed grid while the other processes the spatial point of interest, allowing the output function to be evaluated at any point. Fourier Neural Operator (FNO)~\cite{li2020fourier} parameterizes the transformation in Fourier space, which provides flexibility in the discretization when the Fourier transform and its inverse are computed. The Laplace Neural Operator (LNO)~\cite{cao2024laplace} employs a similar approach to the FNO, but with a Laplace transform to improve performance on nonperiodic problems.}

\textbf{Learning with hard constraints.} Recent efforts in machine learning have explored hard-constrained neural networks as a means to enforce physical or structural constraints on model predictions. These methods can guarantee satisfaction of linear equality constraints \citep{kkt-hpinn, balestriero2023police} or combinations of linear equality and inequality constraints \citep{min2024hard, flores2025enforcing}. Often, these approaches employ either a specific parameterization of a given network or a network-agnostic, closed-form projection layer that maps neural network outputs to a feasible set. Such constraints have been successfully applied to enforce physical constraints such as conservation laws \citep{kkt-hpinn} as well as stability objectives in learned dynamics \citep{min2023data,richards2023learning}. However, these methods are limited to affine constraints. \cite{tordesillas2023rayen} introduce a network parameterization method to enforce certain convex constraints, but the method is limited to input-independent constraints. Other methods \citep{lastrucci2025enforce,grontas2025pinet,goertzen2026hardnet++,tang2026lmi} use iterative approaches to push model outputs towards satisfaction of nonlinear constraints but may not converge under finite iterations. There remains a gap in the existing literature for closed-form guaranteed satisfaction of nonlinear constraints, which are of particular interest for chaotic systems, as enforcing boundedness in predictions can be useful for avoiding trajectory blow-up. 

\textbf{Our contributions.} We introduce the Energy-Constrained Operator (ECO), a framework for learning chaotic dynamics that guarantees long-term stability by construction. By integrating control-theoretic principles, ECO simultaneously learns the dynamics and a stabilizing energy function. To our knowledge, this is the first work to provide explicit formal guarantees on the boundedness of predicted trajectories for such systems. Our key contributions include:
\begin{itemize}
    \item \textbf{A Learnable Energy-Based Constraint:} We derive computationally efficient, algebraic dissipative conditions based on a learnable quadratic Lyapunov function. This allows the model to discover a stabilizing invariant set from data without requiring prior knowledge of the system, which is an outer-estimate for the attractor.
    \item \textbf{A Convex Quadratic Projection Layer:} We design a projection layer that enforces a general convex quadratic constraint, including the dissipativity condition. The layer is network-agnostic, computationally efficient, and fully differentiable.
    \item \textbf{Theoretical Boundedness Guarantees:} We provide a formal proof that our learned model is globally asymptotically stable, ensuring all prediction trajectories remain bounded and converge to the learned invariant set.
    \item \textbf{Empirical Results:} We validate ECO on challenging chaotic benchmarks, including the Kuramoto--Sivashinsky and Navier--Stokes equations. Our model produces stable long-horizon forecasts that accurately reconstruct statistical properties of the strange attractors.
\end{itemize}
\revised{A short version of this paper that contains preliminary results was presented at the 2026 Learning for Dynamics and Control (L4DC) conference \cite{pmlr-v331-goertzen26a}. The current manuscript provides complete proofs for all theoretical results, which were omitted from \cite{pmlr-v331-goertzen26a} due to space constraints. Additionally, we extend the fixed grid energy function definition in \cite{pmlr-v331-goertzen26a} to a general functional formulation, which preserves the discretization-agnostic property of neural operators while maintaining theoretical guarantees. Furthermore, we present extended experimental results, including a Reynolds 500 turbulent flow system, a more challenging counterpart to the Reynolds 40 system in \cite{pmlr-v331-goertzen26a}. Finally, we present additional numerical experiments that demonstrate boundedness guarantees and improved statistical predictions of our method, compared to unconstrained models, across different benchmark systems, backbone model choices, and inference discretizations.\footnote{The codebase for implementing the numerical experiments is available at: \url{https://github.com/azizanlab/ECO}.}}

\section{Problem Formulation}
Consider a chaotic dynamical system described with a PDE of the form,
\begin{equation}\label{eq:continuousdynamics}
\begin{aligned}
    \partial_t{w} = F\left({x},{w},\partial_{{x}}{w},\partial_{{xx}}{w},...\right), \quad\quad &(t,{x})\in[0,T]\times\mathbb{X}\\ 
    {w}(0,{x}) = {w}_0({x}),\quad\quad &{x}\in\mathbb{X}\\
    B\left[{w}\right](t,{x}) = 0, \quad\quad &(t,{x})\in[0,T]\times\partial\mathbb{X}
\end{aligned}
\end{equation}

Here, ${w}(t,{x})$ represents the state of the dynamical system at any time $t\in[0,T]$ and position ${x}\in\mathbb{X}\subseteq \mathbb{R}^d$, ${w}_0({x})$ is the initial condition defined on the full spatial domain ${x}\in\mathbb{X}$, and $B\left[{w}\right](t,{x})$ is the boundary condition defined on the spatial boundary $\partial\mathbb{X}$. We adopt a discrete-time formulation of this problem, and consider PDEs with solutions in $L^2$ space, i.e., $w(t,\cdot) \in L^2(\mathbb{X})$,
\begin{equation}\label{eq:discretedynamics}
    {w}_{t+1}({x}) = G({w}_t(\cdot))({x}),\quad\quad (t,{x})\in\left\{0,1,2,...N\right\}\times\mathbb{X}
\end{equation}
where ${w}_{t}(\cdot) = w(t,\cdot):\mathbb{X}\rightarrow\mathbb{R}$ represents the state of the dynamical system at any position ${x}\in\mathbb{X}$ at time step $t$. We focus on PDEs that govern dissipative chaotic systems, as many physically relevant chaotic systems inherently exhibit dissipative behavior. 


The primary goal is to learn a neural operator $G^*(\theta)$, parameterized by $\theta$, that emulates the true dynamics $G$. For the chaotic systems we study, long-term pointwise prediction is intractable due to sensitive dependence on initial conditions. While individual trajectories are unpredictable, the trajectories of a dissipative chaotic system converge to a statistically invariant strange attractor. Consequently, a more feasible and meaningful objective is to learn an operator that preserves the statistical properties of the true dynamics over long horizons.

A significant challenge to achieving this objective is model stability. When an approximate operator $G^*(\theta)$ is applied iteratively to generate long-horizon trajectory forecasts, small prediction errors can be amplified exponentially, which accumulates and eventually causes predicted trajectories to diverge to unbounded values. This failure mode, also known as ``finite-time blow-up'', prevents the model from capturing any meaningful long-term behavior \citep{lu2018attractor}. 

To address this issue, our solution is to construct a model architecture that is dissipative by design. Since dissipativity is a fundamental property of the physical system, incorporating it as an inductive bias ensures trajectory boundedness without sacrificing expressivity. To achieve this rigorously, we first formalize the concept:

\begin{definition}\label{def:dissipativity}
    We say that the system in ~\eqref{eq:discretedynamics} is \textbf{dissipative} if there exists a bounded (with respect to $L^2$ norm) and positively invariant set $M \subset L^2$ such that $\lim_{t\to\infty} \text{dist}({w}_t, M) = 0, \text{ where } \text{dist}({w}_t, M) = \inf_{y\in M} \|{w}_t - {y}\|$.
    In other words, every trajectory of the system will converge to $M$ asymptotically, and stays within $M$ once it enters. $M$ is said to be \textbf{globally asymptotically stable}.
\end{definition}

Intuitively, a dissipative system loses energy until its trajectories enter and remain within a bounded region $M$, which for chaotic PDEs is their strange attractor. However, Definition~\ref{def:dissipativity} is \emph{descriptive} in the desired property of $M$, without providing a practical mechanism to verify or enforce it. Furthermore, the strange attractor $M$ itself is highly system-dependent and has been known to be computationally intractable to characterize \citep{stuart1998dynamical, milnor1985concept}.

\section{Dissipative Dynamics: A Control-theoretic Perspective}\label{sec:theory}

Our objective is to design a neural operator architecture that enforces dissipativity by construction. The first challenge is that Definition~\ref{def:dissipativity} is non-constructive, as it relies on the strange attractor $M$, which is computationally intractable to characterize. Therefore, we must establish an alternative, computationally efficient condition that can be directly enforced during training and inference.

For this purpose, we turn to a control-theoretic perspective and the concept of Lyapunov functions. These ``energy-like'' functions are extensively used to establish asymptotic stability of dynamical systems \citep{khalil2002nonlinear} and naturally connect with the behavior of dissipative systems. Instead of analyzing the complex attractor $M$, we can use a level set of a Lyapunov function as a tractable proxy for the bounded region.

We adapt this strategy in the following proposition, which generalizes the concept of asymptotic stability from a single equilibrium point to an entire level set, providing the practical conditions needed for our model design.

\begin{proposition}[set asymptotic stability]\label{prop:attractivity}
     For an infinite-dimensional dynamical system in \eqref{eq:discretedynamics}, suppose there exists a non-negative-valued continuously differentiable functional $V: L^2 \to \mathbb{R}_+$ and a constant $c > 0$, such that 
\begin{equation*}\label{eq:dissconditions}
    \begin{aligned}
       (i)&\quad \forall {w}_t \notin M(c)=\{w\in L^2: V({w}) \leq c\}, V({w}_{t+1}) \leq \alpha V({w}_{t}), \\ & \quad0 < \alpha < 1 \\
        (ii)& \quad \forall {w}_t \in M(c)=\{w\in L^2: V({w}) \leq c\}, V({w}_{t+1}) \leq c\\
        (iii)&\quad V \text{ is radially unbounded, i.e., } V(w) \to \infty \text{ as } \|w\| \to \infty
    \end{aligned}
    \end{equation*}
    Then the system \eqref{eq:discretedynamics} is dissipative, where the level set $M(c)$ is globally asymptotically stable.
\end{proposition}
\begin{proof}[Proof for Proposition~\ref{prop:attractivity}]
    By definition, condition (ii) implies that $M(c)$ is indeed a positively invariant set. Since $V$ is radially unbounded, for any $\alpha > 0$, we can find $r_{\alpha}$ such that $V(w) > \alpha$ for all $\|w\| > r_{\alpha}$. Therefore, any level set of $V$ is bounded as $\{x: V(w) \leq \alpha\} \subset B(r_{\alpha})$. Thus, $M(c)$ is both positively invariant and bounded.
    
    Based on the positive invariance property, any trajectory starting with $w_0 \in M(c)$ will always stay within $M(c)$, meaning that $\lim_{t\to \infty} \text{dist}(w_t, M(c)) = 0$ is satisfied.

    Consider a trajectory $\{w_t'\}_{t\in \mathbb{N}}$ that starts outside of the level set, i.e. $w_0' \notin M(c)$ and $V(w_0') > c > 0$. Suppose the trajectory never enters $M(c)$, i.e., $\forall t \in \mathbb{N}, V(w_t') > c$. Using condition (i), we have $V(w_{t+1}') \leq \alpha V(w_t')$, which implies $V(w_t') \leq \alpha^t V(w_0')$. For any $t \geq \log_{\alpha}\left( \frac{c}{V(w_0')}\right)$, $V(w_t') \leq c$ which contradicts the prior assumption. In fact, the trajectory $\{w_t'\}_{t\in \mathbb{N}}$ will enter $M(c)$ in finite time at most $\log_{\alpha}\left( \frac{c}{V(w_0')}\right)$ steps.
\end{proof}

\begin{figure}[h!]
    \centering
    \includegraphics[width=0.8\linewidth]{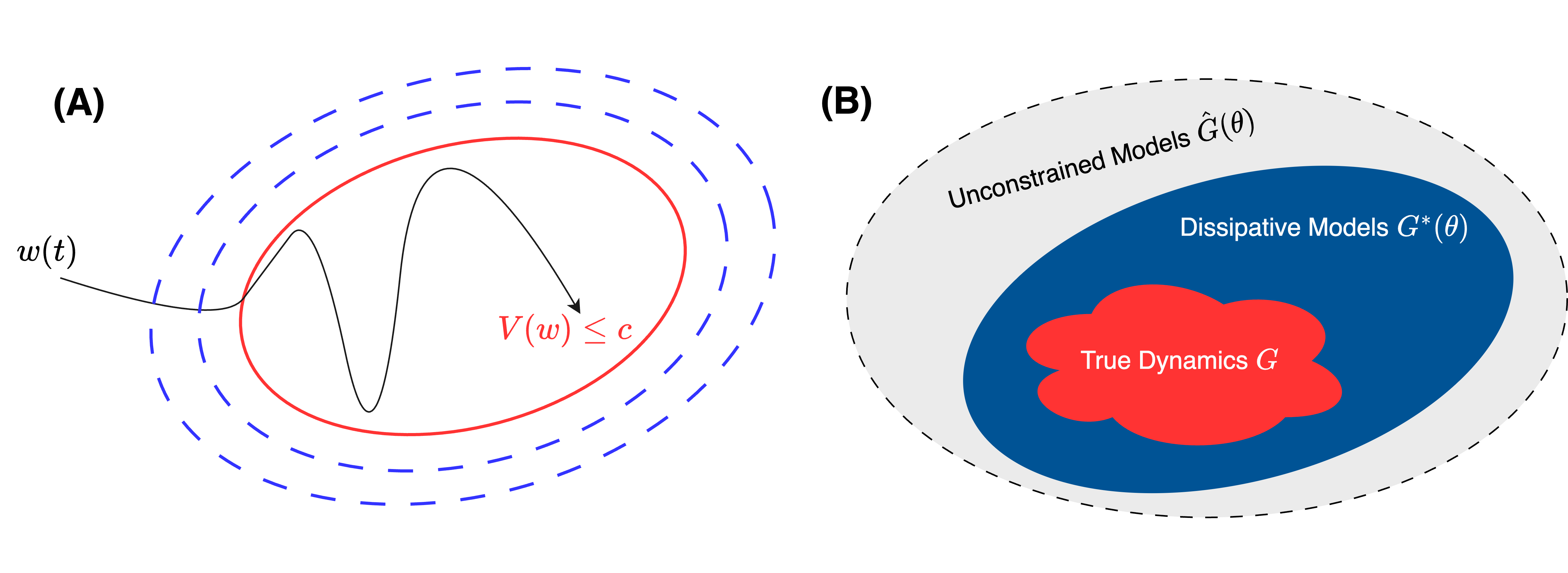}
    \caption{(A) An illustration of the conditions in Proposition~\ref{prop:attractivity}, where the trajectory loses energy over time and enters an invariant level set. (B) Illustration of the fact that an inherently dissipative model would have an effectively smaller search space for parameters.}
    \label{fig:dissipativity_illustration}
\end{figure}

As illustrated in Figure~\ref{fig:dissipativity_illustration}(A), the conditions in Proposition~\ref{prop:attractivity} guide any solution starting outside the level set $M(c)$ to lose energy exponentially due to the $\alpha$ factor, entering the level set in finite time and remaining inside thereafter.

Despite the simplicity of the algebraic conditions derived in Proposition~\ref{prop:attractivity}, overall the conditions still obtain a form of ``if-else'' condition, which might not be straightforward to enforce in a neural network that requires differentiability for backpropagation. To resolve this issue, we unify conditions (i) and (ii) in the above proposition into the following single inequality constraint, which involves a ReLU activation and the $\alpha \in (0,1)$ used in condition \eqref{eq:ineq}:
\begin{equation}\label{eq:ineq}
    V(w_{t+1})-\alpha\left[ V(w_t) + \text{ReLU}\left(c-V(w_t)\right) \right] \leq 0
\end{equation}

Note that the reformulation in \eqref{eq:ineq} is slightly more conservative than the conditions in Proposition~\ref{prop:attractivity}, as the system is asymptotically stable towards $M(\alpha c)$. In practice, $\alpha$ is usually chosen as a value close to 1. More importantly, this reformulation is sufficient for $M(c)$ to be asymptotically stable as $\alpha \in (0,1)$, and we establish the formal theoretical results in Theorem~\ref{thm:main}.


\section{Methodology}

To achieve the goal of building a neural network prediction model that ensures the trajectory it generates always stays bounded, we now introduce a framework that learns dissipative dynamics by design, based on the control-theoretic conditions derived in Section~\ref{sec:theory}. As illustrated in Figure~\ref{fig:dissipativity_illustration}(B), learning an inherently dissipative prediction model conceptually limits the parameter search to a smaller space that is always aligned with physical properties of the true dynamics. Compared to unconstrained models, which might search over parameters that lead to unstable behaviors, our approach makes the training process more efficient.

Our methodology is built on two key components: 1) a learnable Lyapunov functional $V(w)$ that represents the system's energy, and 2) a custom dissipative projection layer that strictly enforces the constraint in \eqref{eq:ineq}. In addition to the learned model being dissipative, our framework is also able to produce an outer-estimate (the level set $M(c)$) for the complex strange attractor without any prior knowledge of the system's invariant statistics, which is known to be difficult to characterize. In what follows, we discuss the details of our architecture design and training procedure.

\subsection{Architecture Design with Boundedness Guarantees}\label{sec:architecture}

We propose a neural network architecture that simultaneously learns the dynamics operator in Equation \eqref{eq:discretedynamics} and an energy-like Lyapunov functional 
\revised{$V=\int q(x)(w_t(x)-w_c(x))^2dx$, with $q(x)>0\,\,\,\forall x$, }which together guarantee the dissipativity conditions in Proposition \ref{prop:attractivity} through the construction of a dissipative projection layer. \revised{In addition to learning a dynamics emulator to predict $w_t(x)$, our approach also learns $q(x)$ and $w_c(x)$, which can be evaluated at any grid point on the spatial domain to compute $V$.}

\begin{figure*}[h]
    \hspace{-10mm}
    \centering
    \includegraphics[width=1.0\textwidth]
    {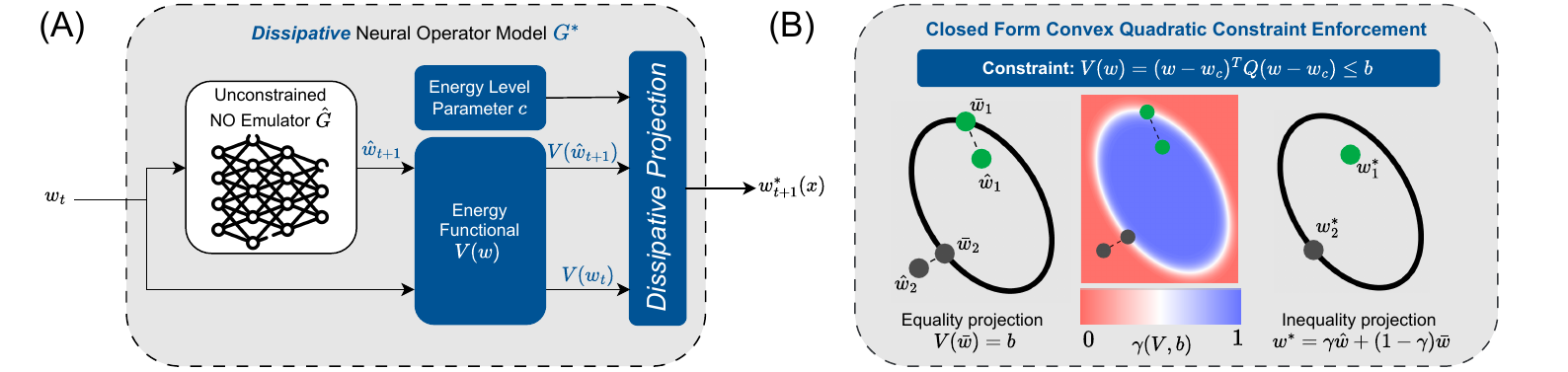}
    \caption{(A) An overview of the proposed model architecture. The input, current time solution $w_t$, is fed into an unconstrained neural operator (NO) emulator $\hat{G}$ to produce a preliminary prediction $\hat{w}_{t+1}$ and a learned energy functional $V$ to compute its energy $V(w_t)$. The dissipative projection layer modifies $\hat{w}_{t+1}$ to produce a final output $w^*_{t+1}$ that satisfies the dissipative energy constraint in \eqref{eq:ineq}. (B) Illustration of the convex quadratic projection for a constraint in the form of $V(w) \leq b$. The equality projection maps any point $w$ to a boundary point $\bar{w}$ in closed form (for both $w_1, w_2$). The quadratic projection is only active for when the constraint is violated, so $w_2$ is projected while $w_1$ is left unchanged. } 
    \label{fig:architecture}
\end{figure*}

Following common practices in learning operators in function spaces \citep{lu2021learning, kovachki2023neural, li2022learning}, 
we consider a discretized spatial domain where the queried spatial location $x \in \mathbb{X}$ is sampled from a finite set $\mathbb{X}_d$ consisting of $n$ grid points, i.e., $x \in \mathbb{X}_d \subset \mathbb{X}$ and the cardinality of $\mathbb{X}_d$ is $n$. As an example, if the spatial domain $\mathbb{X} = [0, 2\pi]$, a fixed grid on $\mathbb{X}$ can be $n$ evenly sampled points, $\mathbb{X}_d = \{k\frac{2\pi}{n-1}: k = 0, 1, ..., n-1\}$. Under the grid setting, the function $w_t \in L^2$ can be effectively represented as an $n$--dimensional vector \revised{$w:= \{w(t, x): x \in \mathbb{X}_d\} \in \mathbb{R}^n$}, which is a collection of solution values at every grid point. \revised{Similarly, the functions $q(x)$ and $w_c(x)$ can also be represented as vectors $\bar{q}\in\mathbb{R}^n$ and $\bar{w}_c\in\mathbb{R}^n$. The energy functional $V$ can then be approximated as a sum $V=\sum_{k=0}^{n-1}q(x_k)\left(w(x_k)-w_c(x_k)\right)^2\Delta x_k$, where $\Delta x_k$ is induced by the choice of discretization. This gives the equivalent form $V=(w-\bar{w}_c)^\top Q(w-\bar{w}_c)$, where $Q\in\mathbb{S}_{++}$ is a positive definite matrix that absorbs both the $\bar{q}(x_k)$ and the $\Delta x_k$ terms.} 
Consequently, the $L^2$ norm of $w_t$ is reduced to a standard 2--norm in $\mathbb{R}^n$. \revised{Importantly, the energy functional inherits the discretization-agnostic trait of neural operators in that any grid points can be used compute $V$, providing flexibility in the choice of discretization during both training and inference.} 
As illustrated in Figure \ref{fig:architecture}(A), our model is composed of two learnable components:
\vspace{-0.5em}
\begin{enumerate}
    \item An unconstrained dynamics emulator $\hat{G}$, which approximates the true dynamics operator $G$ \revised{to predict $w_t(x)$}. The backbone model for the emulator $\hat{G}$ can be any neural operator that maps between function spaces.
    \item A quadratic Lyapunov functional $V(w) = (w - w_c)^T  Q (w - w_c)$, which serves as the energy function. The learnable components \revised{include parameterized $q(x)$ and $w_c(x)$, which, upon discretization, form a} positive definite matrix $Q$ of size $n$--by--$n$ and a center vector $w_c \in \mathbb{R}^n$. 
\end{enumerate}

These components are integrated into a \emph{dissipative projection} layer, which modifies the output of the unconstrained emulator $\hat{G}$ to produce an operator $G^*$ that maps the current solution $w_t$ to the predicted solution at the next time step $w^*_{t+1}$. By construction, the \emph{dissipative projection} layer ensures the condition in \eqref{eq:ineq} is satisfied, which guarantees the predicted dynamical system $w^*_{t+1} = G^*(w_t)$ is dissipative. As a direct consequence of Definition~\ref{def:dissipativity}, all trajectories generated by the operator $G^*$ in an autoregressive manner are guaranteed to be bounded. For a high- dimensional output space, we elect to use a diagonal Q matrix such that the projection can be computed efficiently. 

\subsection{Convex Quadratic Projection Layer}\label{sec:CQP}

The quadratic form of the Lyapunov functional $V(w)$ motivates the development of a convex quadratic projection layer that projects the model predictions onto a feasible set of trajectories where the dynamics are dissipative. We introduce a differentiable convex quadratic projection layer shown in Figure~\ref{fig:architecture}(B) that can handle constraints of the form $(w-w_c)^TQ(w-w_c) \leq b$, where $b$ can be a constant or an arbitrary function of the model input. This form is equivalent to \eqref{eq:ineq}, with $b=\alpha\left[V(w_t)+\text{ReLU}(c-V(w_t))\right]$.

The convex quadratic constraint projection is illustrated in Figure~\ref{fig:architecture}. The general strategy is to define a projection $\bar{w}$ of the model output $\hat{w}$ onto the equality constraint $V(\bar{w})=b$ and selectively project points that violate the constraint $V(\hat{w})\leq b$ onto their respective equality projection. For positive definite $Q$, there exists an explicit form for the projection of $\hat{w}$ onto the equality constraint $V(\bar{w})=b$. That is, for the quadratic Lyapunov function described in Section~\ref{sec:architecture}, the projection 
$\bar{w} = w_c+\sqrt{b}\left(L^T\right)^{-1}\frac{\hat{w} - w_c}{||\hat{w} - w_c||_2}$
satisfies the equation $V(\bar{w}) = b$, where $L$ is the Cholesky decomposition of $Q$ such that $LL^T=Q$. To ensure that this equality projection is only active when the constraint is violated, the final output $w^*$ is calculated as an interpolation between the projected ($\bar{w}$) and non-projected ($\hat{w}$) outputs. 
\begin{equation}\label{eq:quad_proj}
    w^*(x) = \gamma(b,V(\hat{w}))\hat{w}(x) + \left[1-\gamma(b,V(\hat{w}))\right]\bar{w}(x)
\end{equation}
Ideally, $\gamma(b,V(\hat{w}))$ is an indicator function that is $1$ when $V(\hat{w}_t)\leq b$ and $0$ when $V(\hat{w}_t) > b$. However, using an indicator function leads to non-differentiability, which prevents the model from learning a good energy functional $V$ effectively. Instead, we replace the indicator function with sigmoid as a smooth alternative $\gamma(b,V(\hat{w})) = \text{sigmoid}\left[k(b-V(\hat{w}))\right]$. 

\subsection{Theoretical Guarantees}

Our framework is designed to provide formal guarantees of stability and boundedness by construction. These guarantees stem from the convex quadratic projection layer. While this layer utilizes a sigmoid function as a continuous and differentiable relaxation of a strict indicator function, it maintains rigorous theoretical properties. The core idea is that this principled relaxation enforces the constraint on a controllably enlarged invariant set, whose size is governed by the sigmoid's steepness parameter. This is formalized in the following lemma.


\begin{lemma}\label{lem:sigmoid}
    For a quadratic positive definite Lyapunov function, the sigmoid relaxation of the projection in \eqref{eq:quad_proj} maintains the boundedness of the projection output $w^*(x)$, with its energy upper bounded by $V(w^*(x)) \leq (1 + \delta)^2 b$, where $\delta = (2kb+2\sqrt{2kb})^{-1}$ and $k$ is the sigmoid function parameter.
\end{lemma}

The factor $\delta$ provides an explicit, non-asymptotic bound on the relaxation's cost as in the enlargement of the projected ellipsoid for any finite sigmoid function steepness $k$.
\begin{proof}[Proof for Lemma~\ref{lem:sigmoid}]
    Let $\langle u,v\rangle_Q := u^\top Q v$ and $\|u\|_Q := \sqrt{u^\top Q u}$ so that $V(w)=\|w-w_c\|_Q^2$.  
    By construction,
    \[
    \bar w - w_c \;=\; \sqrt b\,(L^\top)^{-1}\frac{\hat w - w_c}{\|\hat w - w_c\|_2}\implies V(\bar w)=\|\bar w-w_c\|_Q^2=b
    \]
    
    Define $\hat{v} := \hat w - w_c$, $\bar{v} := \bar w - w_c$, and $\gamma := \sigma\!\left(k(b - V(\hat w))\right)\in(0,1)$. Then
    \[
    w^* - w_c = \gamma \hat{v} + (1-\gamma)\bar{v}.
    \]
    
    By Minkowski’s inequality for the $Q$-norm,
    \[
    \|w^*-w_c\|_Q
    \;\leq\; \gamma \|\hat{v}\|_Q + (1-\gamma)\|\bar{v}\|_Q
    = \gamma\sqrt{V(\hat w)}+(1-\gamma)\sqrt b.
    \]
    Hence
    \begin{equation}\label{eq:Vw_bound}
        V(w^*) \;\leq\; \big(\sqrt b + \gamma(\sqrt{V(\hat w)}-\sqrt b)\big)^2.
    \end{equation}
    
    If $V(\hat w)\leq b$, then the right-hand side of \eqref{eq:Vw_bound} is at most $b$.  
    So it suffices to consider $V(\hat w)\geq b$.
    
    Let $s=\sqrt{V(\hat w)}\geq\sqrt b$ and $t=s-\sqrt b\geq 0$. Then
    \[
    \gamma = \frac{1}{1+e^{k(s^2-b)}} \;\leq\; \frac{1}{2+k(s^2-b)}
    \]
    since $e^x\geq 1+x$ for all $x\geq 0$.  
    Note that $s^2-b = t(t+2\sqrt b)$, so
    \[
    \gamma t \;\leq\; h(t) := \frac{t}{2+k(t^2+2\sqrt b\,t)}.
    \]
    
    Since the derivative vanishes at $t^\star=\sqrt{2/k}$, let $\delta=(2kb+2\sqrt{2kb})^{-1}$.
    \[
    \begin{aligned}
    h'(t) &= \frac{2-kt^2}{(2+k(t^2+2\sqrt b\,t))^2},
    \implies \\
    \max_{t\geq 0} h(t)
    &= \frac{\sqrt{2/k}}{4+2\sqrt{2kb}}
    = \frac{\sqrt b}{2kb+2\sqrt{2kb}}
    = \delta\sqrt b.
    \end{aligned}
    \]
    
    Thus, for all $s\geq\sqrt b$,
    \begin{equation}\label{eq:gamma_t_bound}
        \gamma(\sqrt{V(\hat w)}-\sqrt b) \;\leq\; \delta \sqrt b.
    \end{equation}
    Combining \eqref{eq:Vw_bound} and \eqref{eq:gamma_t_bound},
    \[
    V(w^*) \;\leq\; \big(\sqrt b + \delta\sqrt b\big)^2
    = (1+\delta)^2 b.
    \]
    
\end{proof}

This guarantee is formalized in the following theorem.

\begin{theorem}\label{thm:main}
    Let the learned dynamics be defined by the operator $w_{t+1}^* = G^*(w_t^*)$, which is composed of an unconstrained neural operator emulator $\hat{G}$ and a dissipative projection layer. Let the learned energy-like function be a quadratic Lyapunov functional $V(w) = (w - w_c)^T Q (w - w_c)$ with learnable center $w_c \in \mathbb{R}^n$ and a positive definite matrix $Q\in \mathbb{S}_{++}^n$. For a choice of $c > 1/\alpha$ and $0 < \alpha < [1 + (2k + 2\sqrt{2k})^{-1}]^{-2}$, the learned dynamical system is dissipative by construction, i.e., the level set $M(c)$ is globally asymptotically stable, and all trajectories generated by the learned dynamics are guaranteed to be bounded.
\end{theorem}
\begin{proof}[Proof for Theorem~\ref{thm:main}]
    Let $b = \alpha[V(w_t) + \text{ReLU}(c - V(w_t))]$, it directly follows from $c>1/\alpha$ that $b \geq \alpha c > 1$. The inequality constraint in \eqref{eq:ineq} states that $V(w_{t+1}) \leq b$. Since $b > 1$, we have $(2kb + 2\sqrt{2kb})^{-1} < (2k + 2\sqrt{2k})^{-1}$. Using Lemma~\ref{lem:sigmoid}, the final output of our model $w^*_{t+1}$ satisfies that $V(w^*_{t+1}) \leq (1 + (2k + 2\sqrt{2k})^{-1})^2 b$. Since $\alpha < [1 + (2k + 2\sqrt{2k})^{-1}]^{-2}$, we can always find some $0 < \alpha_0 < 1$ such that $[1 + (2k + 2\sqrt{2k})^{-1}]^2 \alpha \leq \alpha_0 < 1$. Consequently, the inequality constraint in \eqref{eq:ineq} is satisfied:
    \begin{align*}
        V(w^*_{t+1}) &\leq (1 + (2k + 2\sqrt{2k})^{-1})^2 \alpha [V(w_t) + \text{ReLU}(c - V(w_t))] \\ &\leq \alpha_0 [V(w_t) + \text{ReLU}(c - V(w_t))]
    \end{align*}
    Given the equivalency between the constraint \eqref{eq:ineq} and the conditions in Proposition~\ref{prop:attractivity}, we have now shown the learned dynamics $w^*_{t+1} = G^*(w^*_t)$ is indeed dissipative.
    Note that for the predicted trajectory under the learned dynamics, at any time $t > 0$, $V(w^*_t)$ is strictly bounded by $\text{max}\{V(w_0), c\}$. Let $z = w^*_t - w_c$, the constraint can be rewritten as $z^T Q z \leq \text{max}\{V(w_0), c\}$. Since $Q$ is positive definite, we have $\lambda_{\text{min}}(Q) \|z\|^2 \leq z^T Q z \leq \text{max}\{V(w_0), c\}$. Thus $w^*_t$ will always be bounded with respect to its $L^2$ norm.
\end{proof}

\textbf{Practical Implications.} The condition in Theorem~\ref{thm:main} is non-restrictive and is satisfied using fixed hyperparameters, obviating the need for extensive tuning. For all experiments, we fix the level-set scaling $c\gg 1$ and set $k=100$, which yields the requirement $\alpha<0.9913$. Our chosen value of $\alpha=0.99$ comfortably satisfies this bound. Thus, our framework pairs a rigorous stability guarantee with the flexibility to learn nearly energy-preserving dynamics with a sufficiently large $k$ ($\alpha\to1$ as $k \to \infty$), which is crucial for high-fidelity physical simulations.

\subsection{Training with Invariant Set Volume Regularization}
We construct the training dataset as a collection of $N$ input-output pairs, denoted as $\{(w_i, w_{\text{next}, i})\}_{i=1}^N$, where the input $w_i$ is the current time step solution and the output $w_{\text{next}, i}$ is the next time step solution based on the true dynamics. During training, each input $w_i$ is mapped to a predicted output $w^*_{\text{next}, i}$, and the loss is computed relative to the true next state $w_{\text{next}, i}$. The dynamic loss is defined as the average mean squared error (MSE) between the predicted outputs $\{w^*_{\text{next}, i}\}_{i=1}^N$ and the ground-truth future states $\{w_{\text{next}, i}\}_{i=1}^N$. 

While the convex quadratic projection layer enforces dissipativity and convergence to a level set $M(c)$, it does not inform how to choose an appropriate level set that characterizes the attractor. The goal is to learn the energy functional $V(w)$ such that the resulting ellipsoid is a tight outer estimate of the attractor. To this end, we include a regularization loss in the loss function that penalizes large ellipsoid volumes using Q (defined in Section~\ref{sec:architecture}) and a hyperparameter $\lambda>0$. 
\vspace{-0.5em}
\begin{equation}\label{eq:loss}
    \text{Loss} = \frac{1}{N}\sum_{i=1}^{N} \|w^*_{\text{next}, i}-w_{\text{next}, i}\|_2^2 + \lambda \frac{1}{\sqrt{\text{det}(Q)}}, 
\end{equation}



\section{Numerical Experiments}

\subsection{Methodology illustration via Lorenz 63}
\begin{figure}[h] 
    \centering 
    \subfigure[Long-term trajectory rollout.]{
        \includegraphics[width=0.4\textwidth]{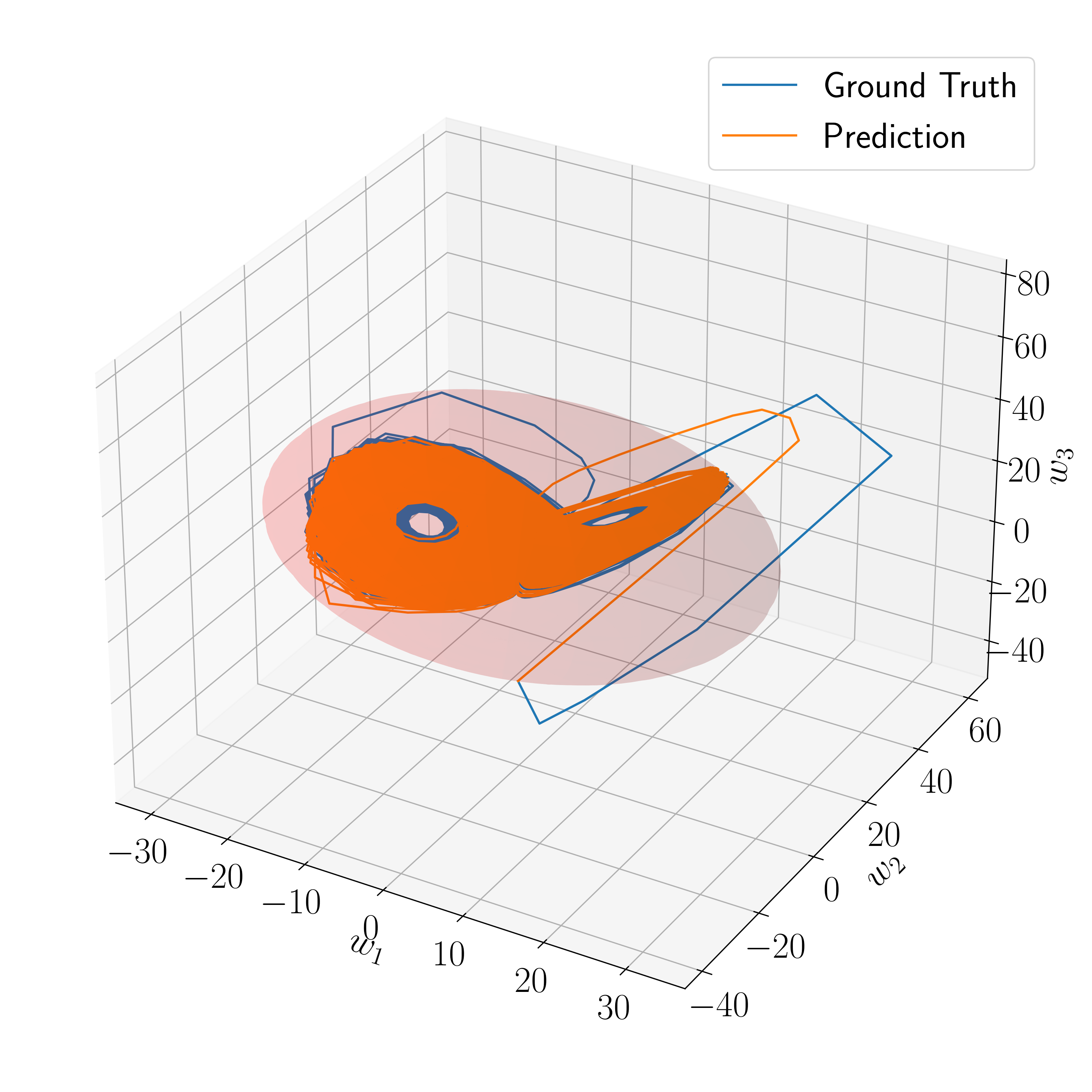}
        \label{fig:L63_rollout}
    }
    \hspace{2em} 
    \subfigure[Learned flow field dynamics.]{
        \includegraphics[width=0.4\textwidth]{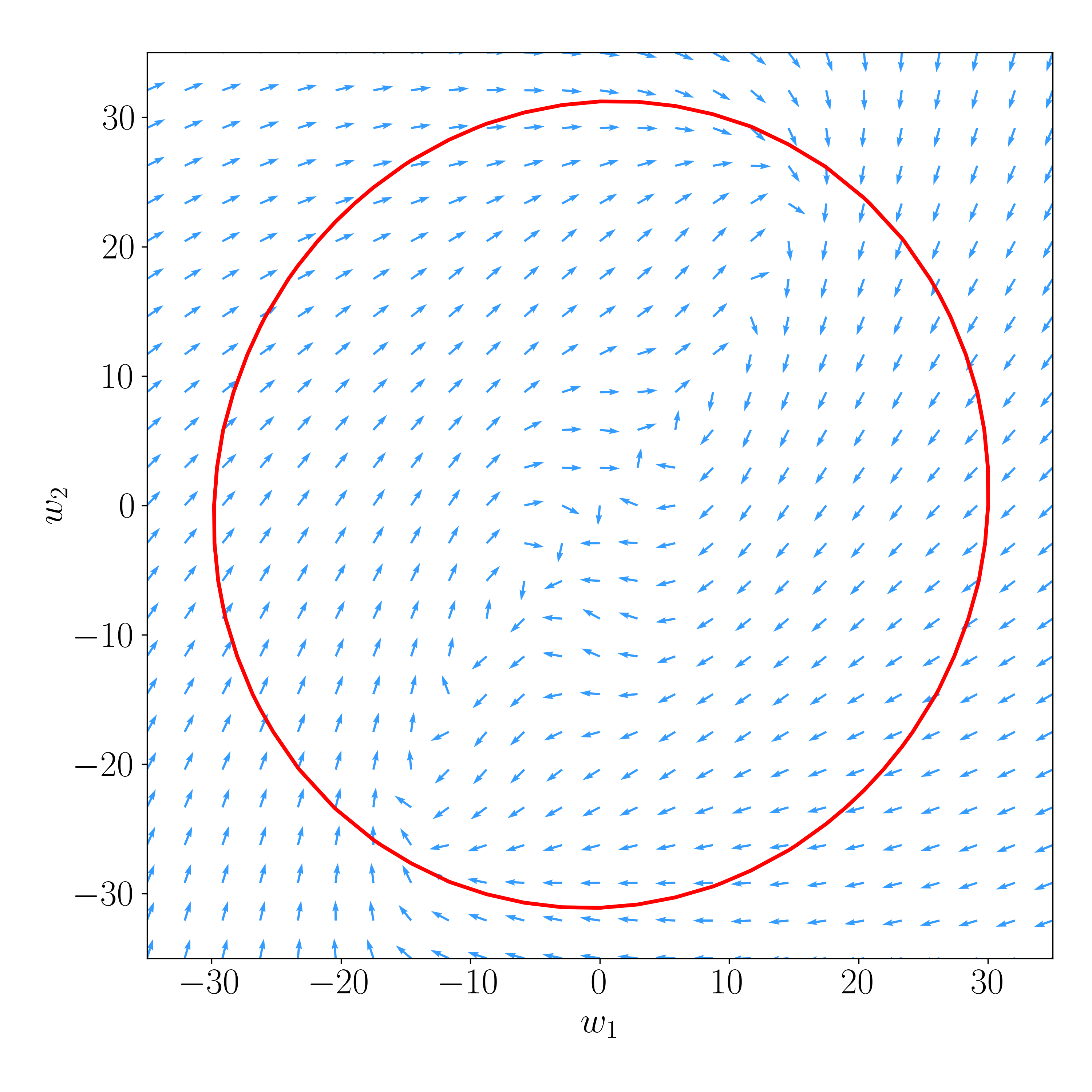}
        \label{fig:L63_flow}
    }
    \vspace{-1em}
    \caption{\textbf{Lorenz 63 prediction results.} (a) A 40,000-step trajectory generated by the model from an unseen initial condition (orange) compared to the ground-truth attractor (blue). The learned invariant ellipsoid (red) is a tight outer-estimate of the strange attractor. (b) The learned flow field on the $w_1$-$w_2$ plane, showing that vectors correctly point inwards across the ellipsoid boundary }
\end{figure}

We first illustrate our methodology on the Lorenz 63 system \citep{lorenz1963deterministic}, a classic low-dimensional model for visualizing chaotic dynamics and strange attractors. As shown in Figure~\ref{fig:L63_rollout}, our model generates a stable, 40,000-step trajectory from an unseen initial condition that accurately recovers the geometry of the true attractor. The learned ellipsoid provides a tight outer-estimate of this attractor, validating the effectiveness of volume regularization in learning a meaningful invariant set.

Figure~\ref{fig:L63_flow} visualizes the dissipative behavior of the learned Lorenz emulator by showing the learned flow field on the $w_1-w_2$ plane. The vector fields along the boundary of the learned ellipsoid all point inwards, ensuring that the set is positively invariant. Together, these results show that our framework successfully enforces long-term stability while learning an energy function that characterizes the system's attractor.

\subsection{\resultedit{Experimental setup, backbone independence, and discretization flexibility}}

\resultedit{We now consider more complex systems governed by chaotic PDEs. Through experiments across different neural operator backbones and various inference resolutions, we demonstrate our method's effectiveness in establishing boundedness guarantees and preserving statistical properties.}

\resultedit{\textbf{Experiment overview.} We validate our framework on the one-dimensional Kuramoto--Sivashinsky (KS) equation \citep{kuramoto1978diffusion} and the two-dimensional Navier--Stokes equations with Kolmogorov forcing \citep{temam2024navier}, both of which are common chaotic PDE benchmarks. \revised{For Navier--Stokes equations, we report results for Reynolds numbers of 40 and 500, where increasing Reynolds number is associated with increased turbulent or chaotic behavior.} 
}

\resultedit{Table~\ref{tab:results} summarizes all experiment configurations and results. For each benchmark system and their backbone neural operator choice, models are evaluated on both the base discretization grid same as the training resolution and a finer discretization grid. In every configuration, we compare the long-term statistical properties obtained by \acro\ against its unconstrained counterpart and the ground truth.}

\resultedit{\textbf{Backbone-agnostic comparisons.}} \resultedit{As discussed earlier in Section~\ref{sec:architecture}, our approach can be applied to an arbitrary backbone neural operator architecture, therefore, it is backbone agnostic. We demonstrate \acro's backbone flexibility by adding our dissipative projection layer to both DeepONet and FNO backbones in experiments. Specifically, in Table~\ref{tab:results}, we present ablation study results which adopts both DeepONet and FNO for the same system, where \acro\ remains bounded and improves upon its unconstrained baseline regardless of the backbone choice. }

\resultedit{\textbf{Discretization-agnostic predictions.}} \resultedit{Due to the learned energy functional design discussed in Section~\ref{sec:architecture}, our approach retains the discretization agnostic property that is central to neural operators such as DeepONet and FNO. In Table~\ref{tab:results}, we present evaluation results to validate the discretization flexibility. Specifically, the models for KS are trained with a 512-point grid and evaluated on a 1024-point grid; the models for NS were trained under a $64\times 64$ grid while evaluated on a $128\times 128$ grid. Throughout these experiments, \acro\ remains bounded and improves statistical predictions on training grids and finer inference grids.}


\resultedit{Detailed discussion of the evaluation metrics, and implementation details including dataset generation and model architecture are included in the Appendix.}

\subsection{\resultedit{Energy Constraints Guarantee Bounded Prediction}}

\resultedit{We now present numerical results demonstrating that \acro's energy constraints ensure boundedness of long-term trajectory predictions across these benchmarks, which illustrates the guarantees established in Theorem~\ref{thm:main}.}

\begin{figure}[tp]
    \centering
    \subfigure[Spatio-temporal plots of trajectory rollout.]{%
        \includegraphics[width=0.6\textwidth]{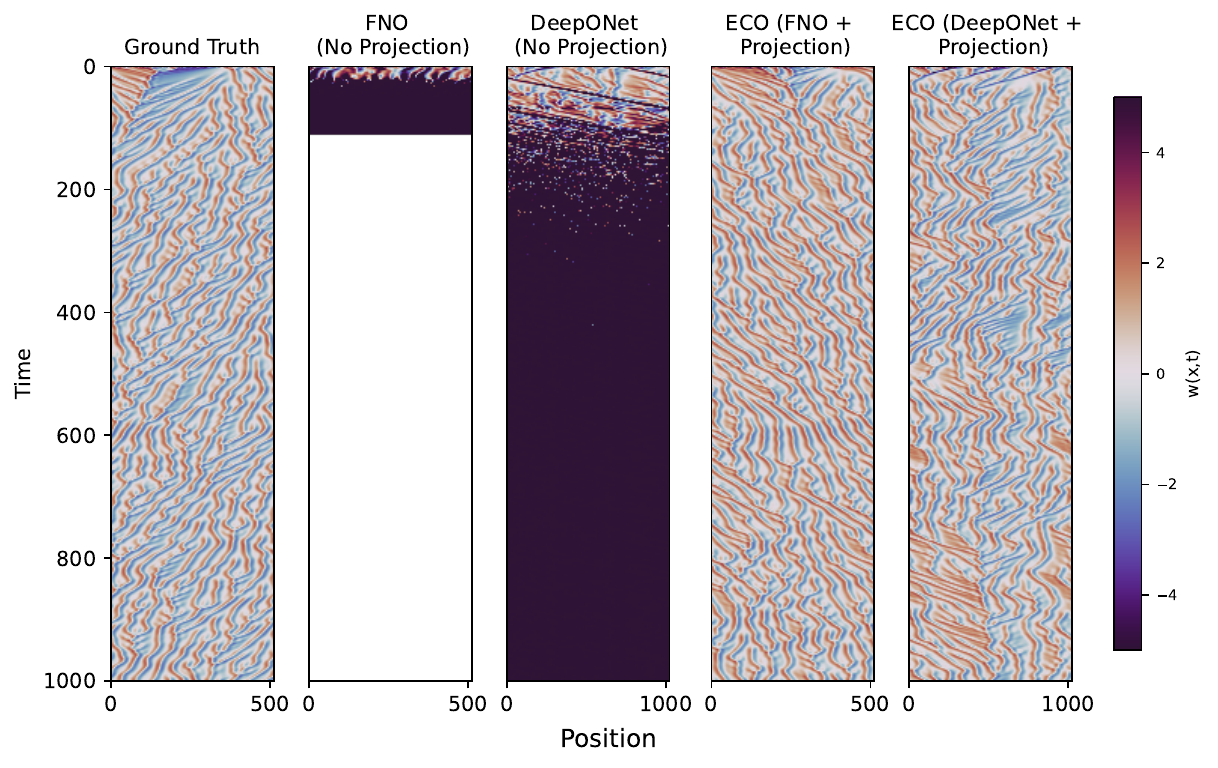}%
    }\\
    \subfigure[Projection of trajectories onto the first two PCA modes.]{%
        \includegraphics[width=0.6\textwidth]{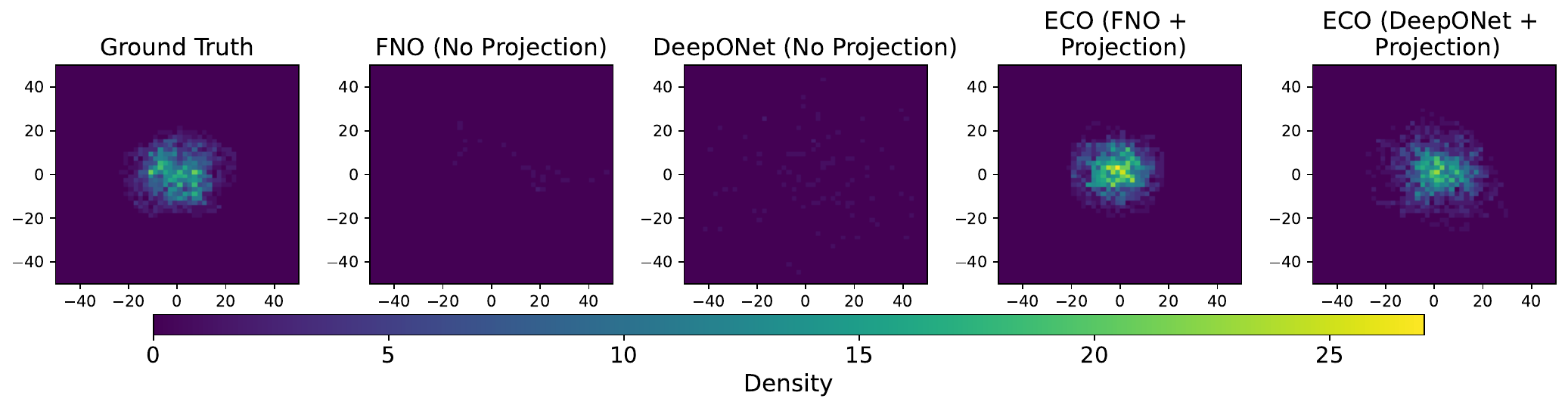}%
    }\label{fig:KS_PCA}\\
    \subfigure[Energy plot of the trajectory rollout.]{%
        \includegraphics[width=0.28\textwidth]{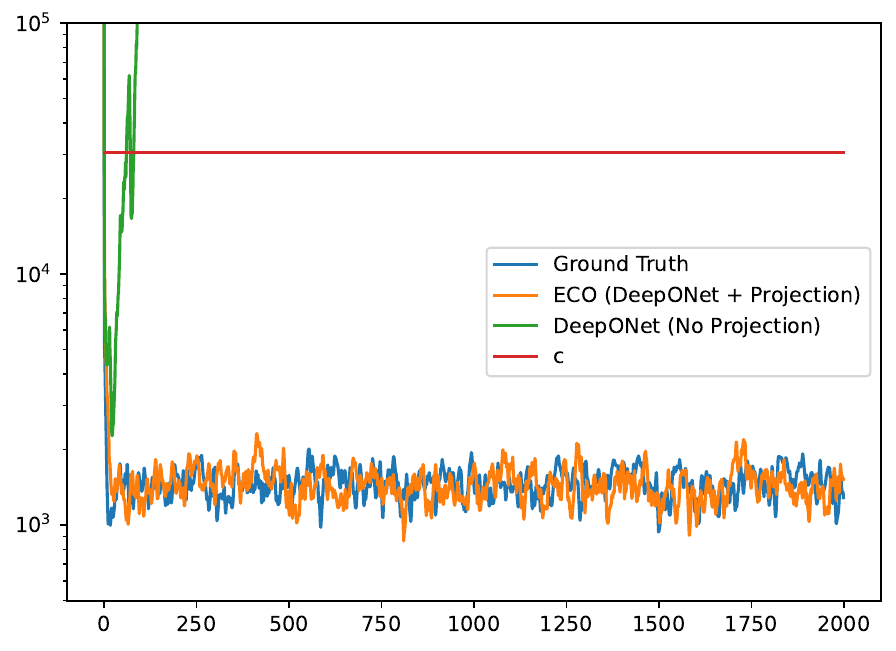}%
    }
    \hspace{1em}
    \subfigure[Fourier energy spectrum of trajectories.]{%
        \includegraphics[width=0.28\textwidth]{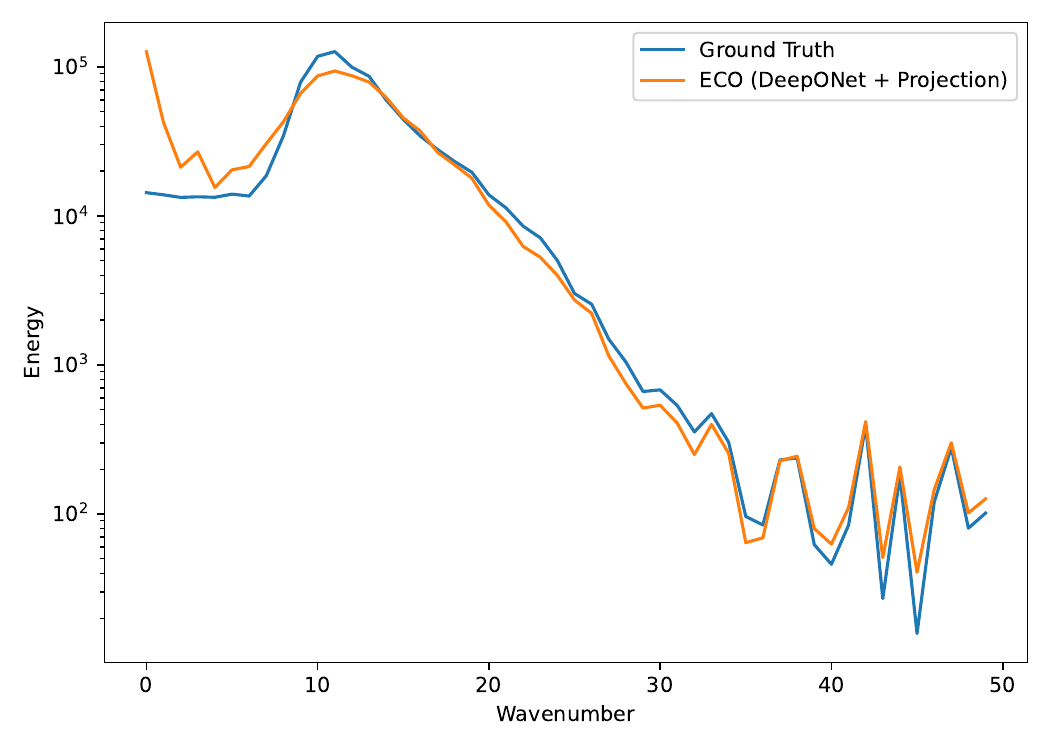}%
    }
    \caption{\textbf{KS prediction results for constrained (ECO) and unconstrained models} (a) Comparison of ground-truth trajectory with predictions from vanilla and projected model. Trajectories are visualized for 1000 seconds after a 500-second transition. The vanilla model blows up, while the projected model stays bounded. The FNO results are presented on a 512 grid and the DeepONet results are presented on a 1024 grid as an example of the discretization flexibility of our approach. (b) Projections of the trajectories onto the first two PCA modes. The projected models follow a similar distribution to the ground truth. (c) Energy plot for an unconstrained DeepONet and \acro\ against the ground truth energy. The energy of the unconstrained model grows, while the energy of \acro\ remains bounded and closely matches the ground truth. (d) The energy spectrum of \acro\ closely follows that of the ground truth}
    \label{fig:KS_results}
\end{figure}

\resultedit{\textbf{Kuramoto--Sivashinsky.}} The comparisons in Figure~\ref{fig:KS_results} highlight the critical role of the projection layer. When rolled out for 2000 time steps from an unseen initial condition, \revised{both the unconstrained FNO and unconstrained DeepONet models} quickly become unstable, and their predictions blow up. In contrast, our projected model remains bounded and successfully recovers the complex spatio-temporal flow patterns of the true dynamics (Figure~\ref{fig:KS_results}a). \revised{Our projection layer keeps predictions bounded, allowing the expressive power of the backbone to reconstruct the dynamics. Figure~\ref{fig:KS_results}c demonstrates that as the dynamics are rolled out, the unconstrained model quickly reaches large energy values, while the energy of the \acro\ model remains bounded and closely matches that of the ground truth.}

\resultedit{It is also noteworthy that in Figure~\ref{fig:KS_results}, we present the comparison results over two common neural operator backbones, DeepONet and FNO. In both cases, our model (\acro) prediction remains bounded and recovers the spatio-temporal patterns of the ground truth while its unconstrained counterpart experiences finite-time blowup, which validates the backbone flexibility of our approach.}

\begin{figure}[htbp] 
    \centering 
    \subfigure[Snapshots of NS flow behavior]{
        \includegraphics[width=0.6\textwidth]{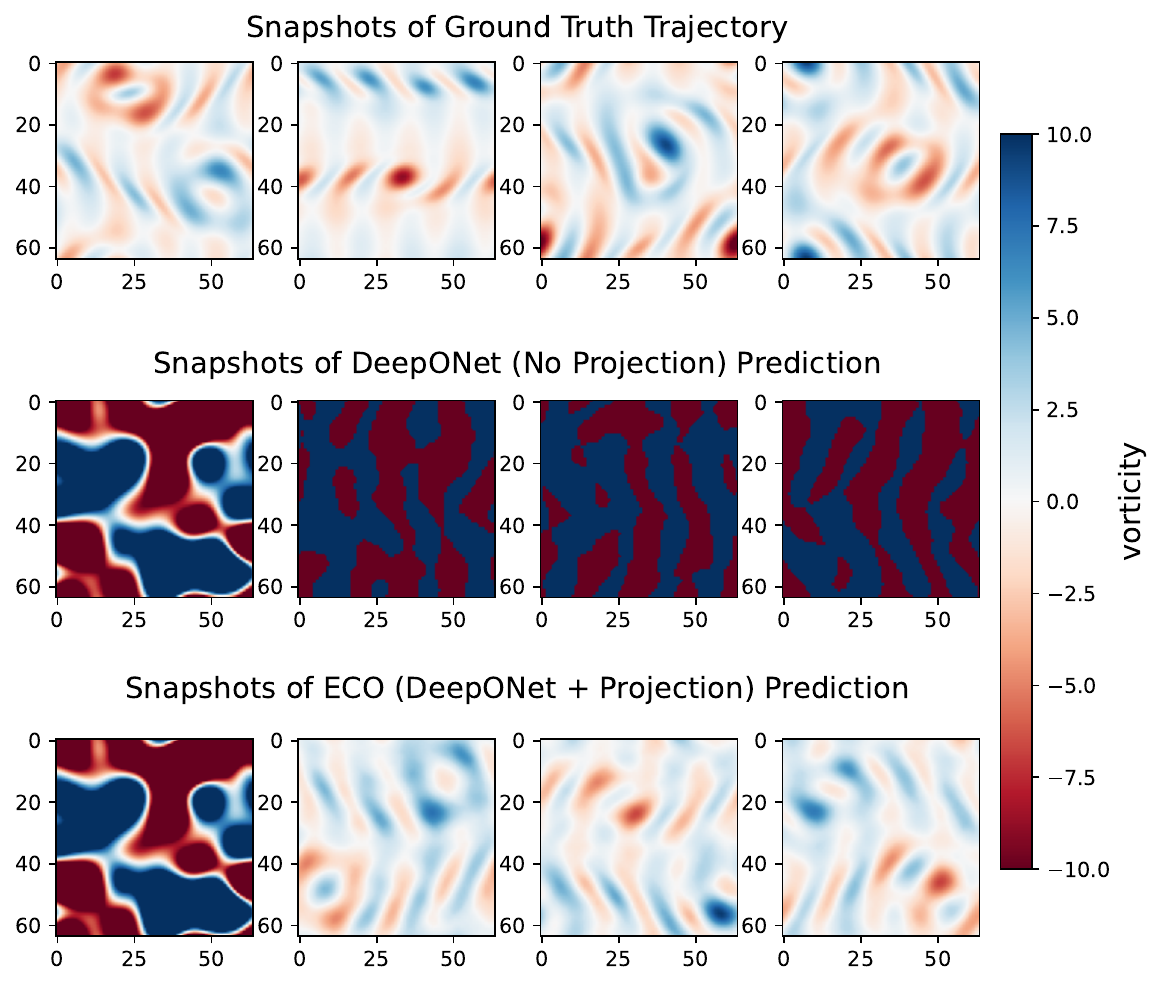}
        \label{fig:NS40_snapshots}
    }
    \\ 
    \subfigure[Projection of trajectories onto PCA modes.]{
        \includegraphics[width=0.5\textwidth]{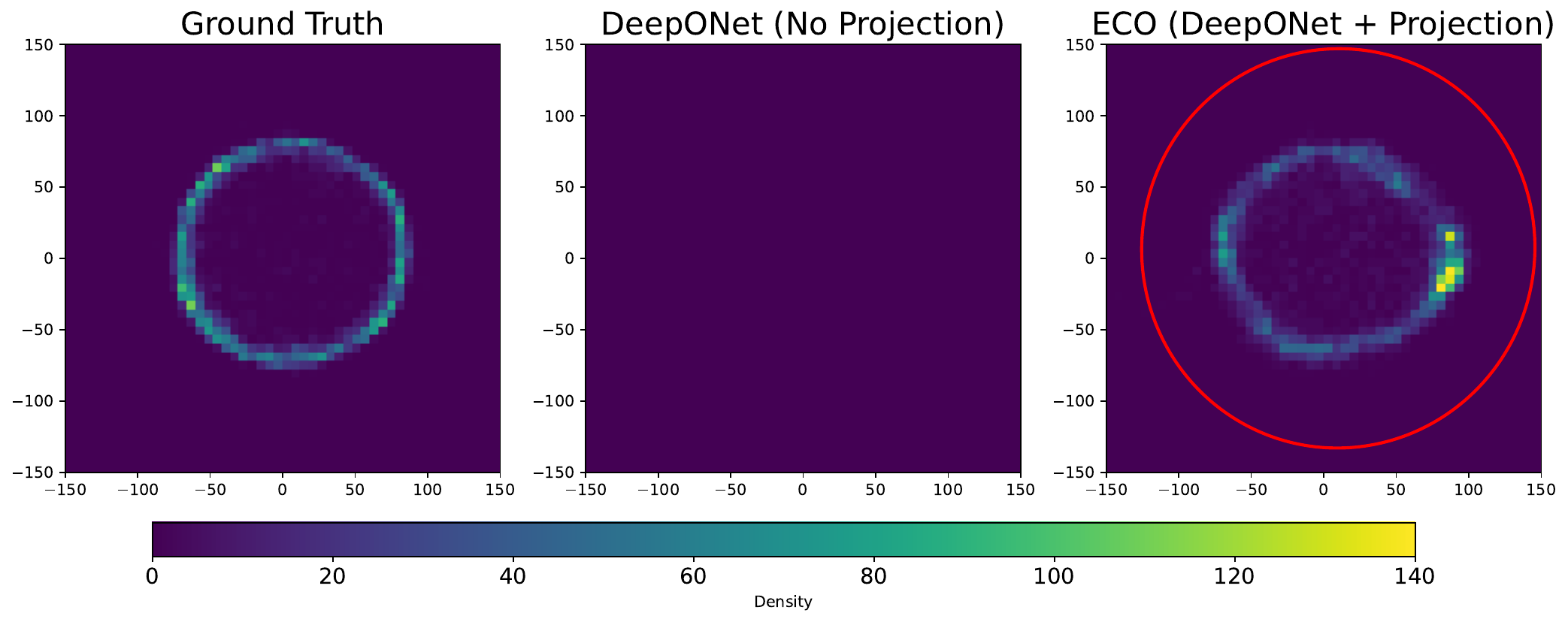}
        \label{fig:NS_PCA}
    }
    \caption{\textbf{NS prediction results.} (a) Snapshots of the flow at various time steps for the model-predicted and ground-truth Navier--Stokes trajectories. The ground-truth and model-predicted dynamics exhibit similar patterns. (b) Projection of trajectories onto the first two PCA modes. The learned ellipsoid forms a tight outer bound on the attractor.}
    \label{fig:NS_results_Re40}
    \label{fig:NS_DeepONet_Re40}
\end{figure}

\resultedit{\textbf{Navier--Stokes at $Re=40$.}} \revised{In the Reynolds 40 case (Figure~\ref{fig:NS_results_Re40}),} when rolled out autoregressively from an unseen initial condition for 10,000 time steps, the baseline DeepONet quickly grows to large values deviating from the attractor, as shown in Figure~\ref{fig:NS40_snapshots}. In contrast, our model (\acro) ensures trajectory boundedness near the attractor, hence generating a flow profile that closely resembles the patterns seen in the snapshots of the ground-truth trajectory.

\begin{figure*}[h]
    \centering

    \begin{minipage}[c]{0.55\textwidth}
        \centering
        \subfigure[Snapshots of NS flow behavior]{
            \includegraphics[width=\linewidth]{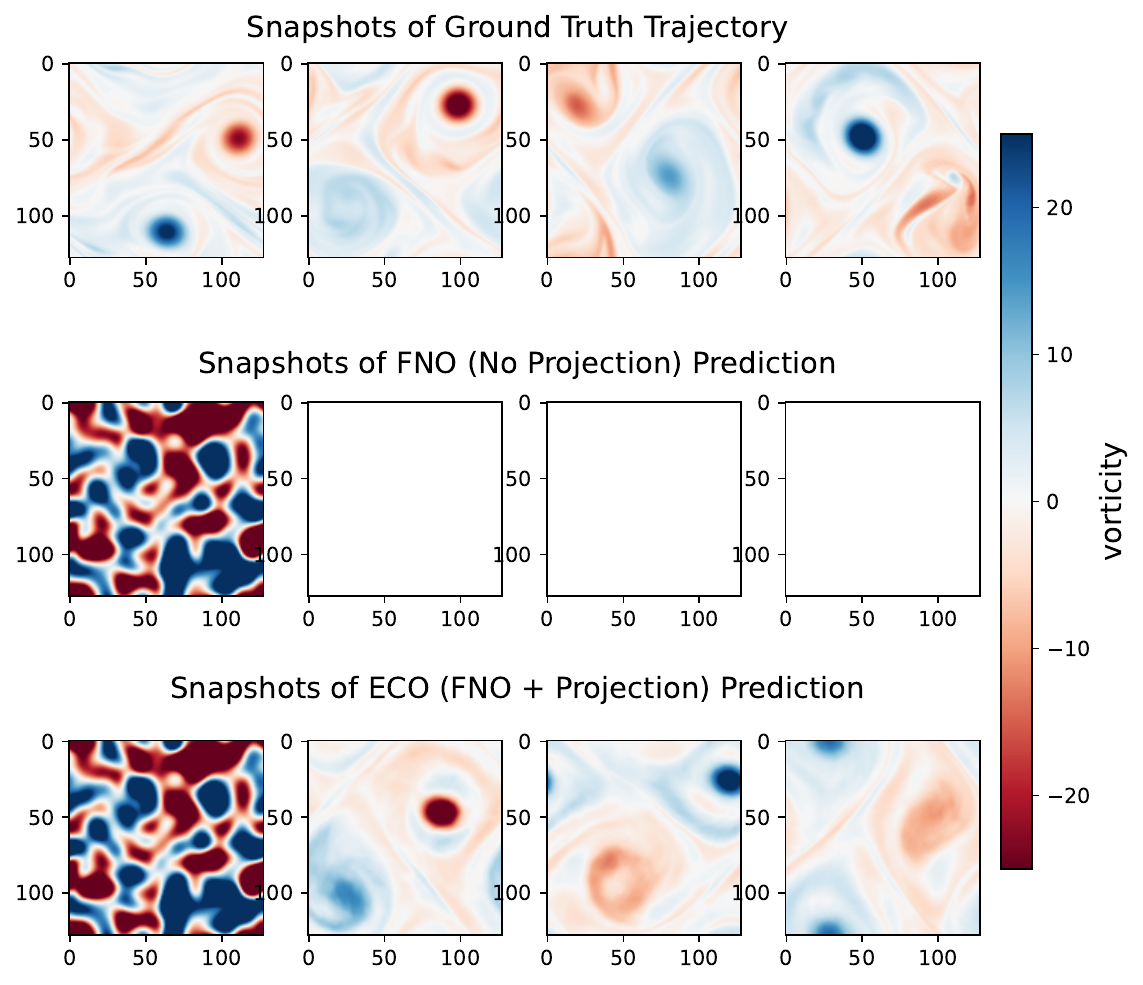}
            \label{fig:NS500_snapshots}
        }
    \end{minipage}
    \hfill
    \begin{minipage}[c]{0.2\textwidth}
        \centering
        \subfigure[Projection of trajectories onto PCA modes.]{
            \includegraphics[width=\linewidth]{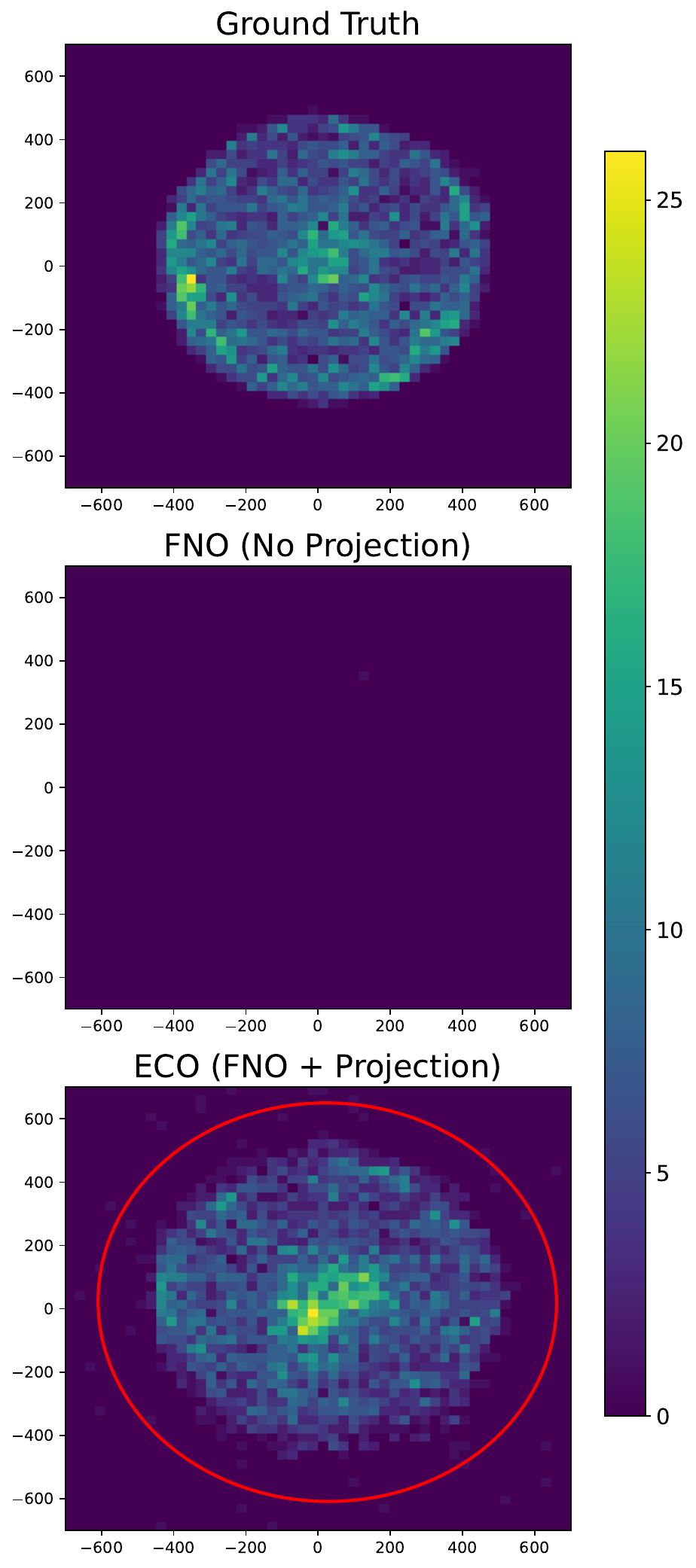}
            \label{fig:NS_PCA_500}
        }
    \end{minipage}
    \hfill
    \begin{minipage}[c]{0.2\textwidth}
        \centering

        \subfigure[Energy plot of trajectory rollout.]{
            \includegraphics[width=\linewidth]{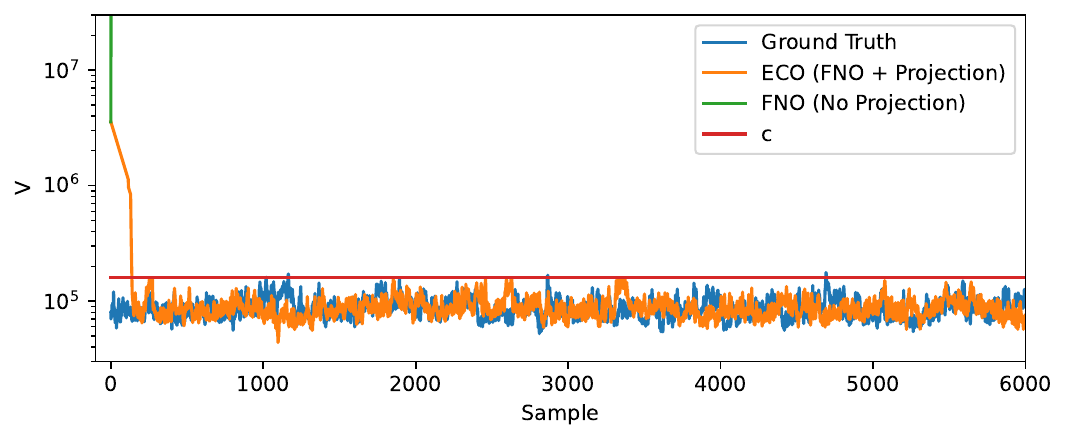}
        }

        \vspace{0.5em}

        \subfigure[Probability density of trajectories]{
            \includegraphics[width=\linewidth]{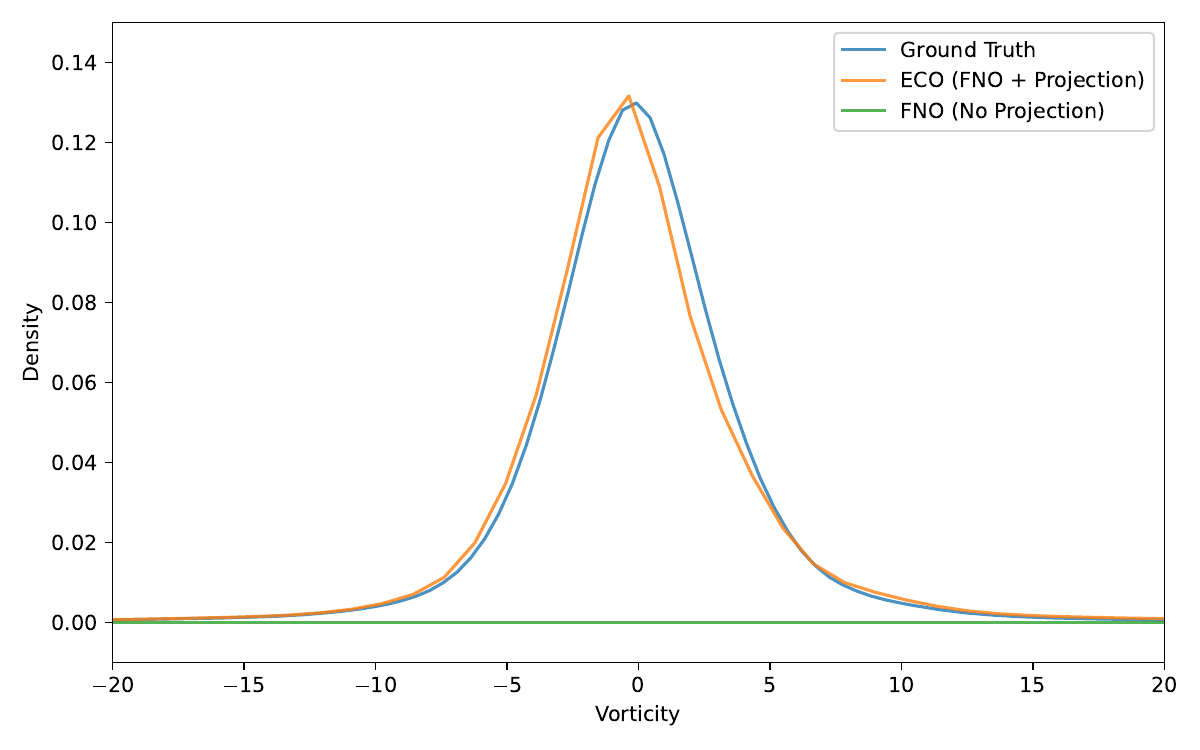}
        }

        \vspace{0.5em}

        \subfigure[Fourier energy spectrum of trajectories.]{
            \includegraphics[width=\linewidth]{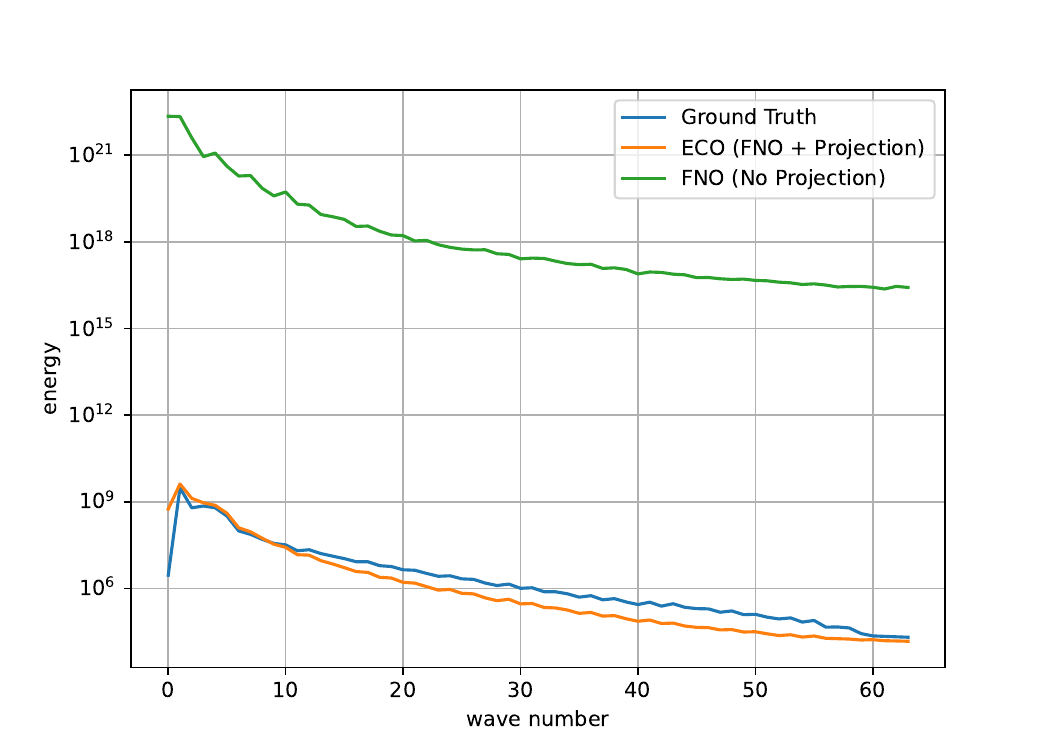}
        }

    \end{minipage}

        \caption{\textbf{NS prediction results (128x128).} (a) Snapshots of the flow at various time steps for the model-predicted and ground-truth Navier--Stokes trajectories. The ground-truth and model-predicted dynamics exhibit similar patterns. (b) Projection of trajectories onto the first two PCA modes. The learned ellipsoid forms a tight outer bound on the attractor. \resultedit{(c) Learned energy along the rollout. (d) Vorticity distributions of the ground-truth and predicted trajectories. (e) Fourier energy spectra of the ground-truth and predicted trajectories.}}
    \label{fig:NS_results_Re500_128}
\end{figure*}

\resultedit{\textbf{Navier--Stokes at $Re=500$.}} \revised{Results for the more chaotic Reynolds 500 case with a constrained (\acro) and unconstrained FNO are shown in Figure~\ref{fig:NS_results_Re500_128}. Trajectories are rolled out autoregressively from an unseen initial condition for 8,000 time steps. While the unconstrained FNO blows up, inhibiting its ability to generate physically realistic predictions, our model (\acro) generates bounded predictions that are physically consistent with the ground truth.}

\revised{This is further illustrated in Figure~\ref{fig:NS_results_Re500_128}(c). The energy plot shows several instances where the energy of the \acro\ trajectory reaches the upper bound on the level set (i.e., instances where the constraint is active), demonstrating the utility of our method in maintaining bounded predictions.}

\subsection{\resultedit{Improvements in Statistical Accuracy}}

To quantitatively validate our approach, we compare the statistical accuracy of {\acro} against the unconstrained baseline on the two PDE benchmarks. As shown in Table~\ref{tab:results}, our method significantly improves key statistical metrics like the KL divergence and log-spectral distance (detailed in the Appendix~\ref{append:metrics}).


\begin{table*}[h!]
\centering
\small
\renewcommand{\arraystretch}{1.15}
\setlength{\tabcolsep}{5pt}

\begin{tabular}{llccc}
\toprule
\textbf{System} &
\textbf{Configuration} &
\textbf{KL Div.} &
\textbf{KL Div.} &
\textbf{Log-Spectral} \\
&
&
\textbf{(physical)} &
\textbf{(PCA)} &
\textbf{Distance} \\
\midrule

\multirow{8}{*}{Kuramoto--Sivashinsky}

& FNO-512, Baseline
& 5.213 & 14.83 & 29.25 \\

& FNO-512, ECO
& \textbf{0.09057} & \textbf{1.000} & \textbf{0.03211} \\

\cmidrule(lr){2-5}

& FNO-1024, Baseline
& 6.943 & 14.20 & 27.76 \\

& FNO-1024, ECO
& \textbf{0.04930} & \textbf{2.141} & \textbf{0.5040} \\

\cmidrule(lr){2-5}

& DeepONet-512, Baseline
& 3.355 & 13.40 & 32.86 \\

& DeepONet-512, ECO
& \textbf{0.02992} & \textbf{1.760} & \textbf{0.1317} \\

\cmidrule(lr){2-5}

& DeepONet-1024, Baseline
& 3.101 & 12.77 & 32.67 \\

& DeepONet-1024, ECO
& \textbf{0.01897} & \textbf{3.116} & \textbf{0.1981} \\

\midrule

\multirow{6}{*}{Navier--Stokes ($Re=40$)}

& FNO-64, Baseline
& \textbf{0.04204} & 0.7574 & 0.6170 \\

& FNO-64, ECO
& 0.04389 & \textbf{0.5677} & \textbf{0.4791} \\

\cmidrule(lr){2-5}

& FNO-128, Baseline
& 0.3608 & 3.889 & 1.082 \\

& FNO-128, ECO
& \textbf{0.04564} & \textbf{0.6916} & \textbf{0.5318} \\

\cmidrule(lr){2-5}

& DeepONet-64, Baseline
& 1.066 & 2.323 & 14.93 \\

& DeepONet-64, ECO
& \textbf{0.002997} & \textbf{1.062} & \textbf{0.8127} \\

\midrule

\multirow{4}{*}{Navier--Stokes ($Re=500$)}

& FNO-64, Baseline
& 1.530 & 14.25 & 12.79 \\

& FNO-64, ECO
& \textbf{0.003528} & \textbf{0.3283} & \textbf{0.4957} \\

\cmidrule(lr){2-5}

& FNO-128, Baseline
& 1.524 & 14.54 & 11.81 \\

& FNO-128, ECO
& \textbf{0.008406} & \textbf{0.2858} & \textbf{0.5342} \\

\bottomrule
\end{tabular}
\caption{\revised{Comparison of baseline models and ECO across all systems, backbones, discretizations, and Reynolds numbers. For a given system, the model was trained on the \textit{lower} discretization (512 for Kuramoto-Sivashinsky and 64x64 for Navier-Stokes) and evaluated on both the training discretization and a higher discretization (512 and 1024 for Kuramoto-Sivashinsky and 64x64 and 128x128 for Navier-Stokes). Lower values are better for all metrics.}}
\label{tab:results}
\end{table*}

\resultedit{\textbf{Kuramoto--Sivashinsky.}} Our method not only ensures boundedness but also successfully recovers the system's invariant statistics. By projecting the trajectories onto their first two principal components, Figure~\ref{fig:KS_results}b shows that our model's predictions exhibit a similar distribution as the ground-truth data, demonstrating our model's capability to reconstruct invariant statistics on the strange attractor. The unconstrained model, prior to divergence, samples a sparse and unstructured distribution. 
\revised{Additionally, Figure~\ref{fig:KS_results}d shows close alignment between the Fourier spectra of the ground truth and \acro\ model for a DeepONet backbone. This is further supported by Table~\ref{tab:results}, which shows that the log-spectral distance between the ground truth and the modeled trajectories is large for the unconstrained model, while remains consistently small for the projected model across all Kuramoto--Sivashinsky experiments.} 

\resultedit{\textbf{Navier--Stokes at $Re=40$.}} Furthermore, the stability allows \acro\ to accurately capture the system's underlying statistical structure. As shown in Figure~\ref{fig:NS_results_Re40}(b), when trajectories are projected onto the first two principal components, our model's predictions correctly reproduce the distinct ring-shaped geometry of the true attractor, while the unconstrained model fails to do so. \revised{This is further quantified in Table~\ref{tab:results}, where we demonstrate that the KL divergence between the PCA distributions of the ground truth and \acro\ remains small for all $Re=40$ experiments.}

\resultedit{\textbf{Navier--Stokes at $Re=500$.}} \revised{Furthermore, the projection of the dynamics and the learned ellipsoid bound onto the first two PCA modes demonstrate that the learned energy functional induces a tight outer bound on the dynamics.}

\revised{Figures~\ref{fig:NS_results_Re500_128}(d) and (e) further demonstrate how enforcing this boundedness leads to accurate statistical reconstruction. The vorticity probability density plot shows a close match in the distributions induced by the ground truth and \acro\ model, while the dynamics predicted by the unconstrained FNO does not generate substantial probability density along the ground truth physical regime. The Fourier spectra also show a close match between the constrained model and the ground truth, while the unconstrained model produces much higher-energy predictions.}

It is noteworthy that these gains are achieved without providing any prior information about the system's dynamics or its attractor, as our model learns the stabilizing energy function and the dynamics purely from data.

\resultedit{\textbf{Discretization-agnostic property.} Also of note is that several presented results are evaluated on discretizations different from the training grids, which validates the discretization-agnostic property retained by our approach. In Figure~\ref{fig:KS_results}, the models for KS were trained using a 512-point grid while being evaluated on a finer 1024-point grid; Similarly, in Figure~\ref{fig:NS_results_Re500_128}, the FNO-based models for NS were trained using a $64\times 64$ grid while being evaluated with $128\times 128$ resolution. As reported in Table~\ref{tab:results}, in cases such as KS and NS $Re=500$, \acro's boundedness guarantees result in significant improvement in statistical accuracy when compared to unconstrained baselines on the finer evaluation grids.}




\section{Conclusion}
To address the fundamental challenge of long-term stability in data-driven chaotic dynamics models, we introduced ECO, an energy-constrained operator design. Our framework jointly learns the system dynamics and a quadratic energy functional, and proposes a novel convex quadratic projection layer to enforce a computationally efficient dissipativity condition. This approach, to our knowledge, is the first to provide explicit theoretical guarantees on the long-term boundedness of a learned chaotic emulator. We demonstrated ECO's effectiveness on benchmarks, including Kuramoto--Sivashinsky and Navier--Stokes, where it produced stable, long-horizon forecasts that accurately reproduced the systems' invariant statistics while unconstrained baselines failed. This work demonstrates the value of building physically constrained models, marking a step toward more reliable and robust scientific machine learning.


\section*{Acknowledgment}
The authors acknowledge the MIT SuperCloud and Lincoln Laboratory Supercomputing Center for providing computing resources that have contributed to the results reported within this paper. This work was supported in part by MathWorks, the MIT-Amazon Science Hub, the MIT-Google Program for Computing Innovation, the MIT-IBM Watson AI Lab, and the National Science Foundation Graduate Research Fellowship under Grant No. 2141064.

\bibliographystyle{unsrtnat} 
\bibliography{ref}

\appendix
\section{Metrics for Statistical Property Evaluation}\label{append:metrics}
We use three statistical metrics to evaluate the effectiveness of our model in capturing statistical properties. 

\textbf{KL Divergence (physical):} We compare the pixel-wise distribution of velocity (Kuramoto--Sivashinsky) or vorticity (Navier--Stokes) between the ground-truth data and the model-predicted trajectories. We evaluate the distance between distributions with the Kullback--Leibler divergence: 
\begin{equation}\label{eq:KLdiv}
    D_{KL}(P||Q)=\sum_{x\in\mathcal{X}} P(x)\text{log}\frac{P(x)}{Q(x)}
\end{equation}
In this case, $x\in\mathcal{X}\subset\mathbb{R}$ is the support of the physical quantity (velocity or vorticity) across all pixels in space and time. That is, a single grid point on the spatial domain at a single snapshot in time represents one sample from the distribution. We use $P(x)$ as the ground-truth distribution and $Q(x)$ as the predicted distribution.

\textbf{KL Divergence (PCA):} We also compare the Kullback--Leibler divergence between the distributions of the physical quantities onto their two PCA modes. Projecting every snapshot in time onto the first two PCA modes gives a 2D distribution of the trajectories in PCA space, as seen in Figures~\ref{fig:KS_results} and~\ref{fig:NS_PCA}. We then compute the KL divergence from Equation~\ref{eq:KLdiv}, where $x\in\mathcal{X}\subset\mathbb{R}^2$ is the support of the projected snapshot onto a 2D point across all snapshots in time. We use $P(x)$ as the ground-truth distribution and $Q(x)$ as the predicted distribution.

\textbf{Log-Spectral Distance:} We create energy spectra for predicted trajectories by performing a discrete Fourier transform in the spatial domain of each snapshot in time. The energy of a given Fourier mode is computed by taking the average in time of the square signal. The Log-Spectral Distance between the ground-truth spectrum and the predicted spectrum is defined as 

\begin{equation}
    D_{LS} = \left(\frac{1}{N}\sum_{n=1}^N\left| \text{log}P(n)-\text{log}\hat{P}(n)\right|^p\right)^{1/p}
\end{equation}
Here, $P(n)$ is the ground-truth energy of the $n$th Fourier mode and $\hat{P}(n)$ is the predicted energy of the $n$th Fourier mode. We use $p=2$ for the results reported in Table~\ref{tab:results}. 

\section{Implementation Details}

\subsection{Lorenz 63}\label{sec:L63_details}

\textbf{Governing Equations.} The Lorenz 63 system is described by the following ordinary differential equations (ODEs):
\begin{align*}
    \dot{w}_1 = \sigma (w_2 - w_1), \
    \dot{w}_2 = w_1(\rho - w_3) - w_2, \ 
    \dot{w}_3 = w_1 w_2 - \beta w_3.
\end{align*}
where $w \in \mathbb{R}^3$ is the system state. We use parameters $\sigma=10.0, \rho=28.0, \beta = 8/3$ which generate chaotic behaviors. 

\textbf{Dataset Generation. } The training data was generated from a single long trajectory integrated for 10,000 seconds with a sampling interval of 0.05 seconds. The trajectory was initialized at a random state outside the strange attractor to ensure the model learns the dissipative flow. For evaluation, we generated a test trajectory by starting from a new, unseen initial condition for 2,000 seconds.

\textbf{Model Architecture. } This system can be viewed as a special case of an infinite-dimensional dynamical system where the spatial domain consists of just three discrete points. In this simplified context, backbone neural operators like the Fourier Neural Operator \citep{kovachki2023neural} and DeepONet \citep{lu2021learning} reduce to a simple multilayer perceptron (MLP). Accordingly, we constructed our unconstrained neural operator emulator, $\hat{G}$, as a feedforward neural network with 6 hidden layers, each containing 150 neurons.

\subsection{Kuramoto--Sivashinsky}\label{sec:KS_results}
\textbf{Governing Equations.} \begin{equation}\label{eq:KS}
\begin{aligned}
w_t + w_{xx}+w_{xxxx}+\frac{1}{2}(w^2)_x = 0,\quad& (t,x)\in[0,\infty)\times[0,L]\\
w(0,x) = w^0(x),\quad& x\in[0,L] 
\end{aligned}
\end{equation}

We use a domain length $L = 32\pi$ to generate chaotic behavior.

\textbf{Dataset Generation. } The training dataset consists of six trajectories simulated for $500$ seconds, with snapshots saved every $1$ second at a spatial resolution of $512$ points. The validation set contains two trajectories with the same discretization. All trajectories were initialized with random conditions.

\textbf{Model Architecture. } 
We construct two models based on DeepONet backbones, one with the added projection layer discussed in Section~\ref{sec:CQP} and one vanilla model without. For both models, the DeepONet branch network consists of three convolutional layers of output dimension 32, 64, 128, and two fully connected layers, each with 256 neurons. The trunk network consists of four fully connected layers, each with 256 neurons.

\subsection{Navier--Stokes}\label{sec:NS_results}
\textbf{Governing Equations. } The 2D Navier--Stokes equations in vorticity form with Kolmogorov forcing are given by:
\begin{equation}\label{eq:NS}
\begin{aligned}
w_t &= - u\cdot\nabla w + \frac{1}{Re}\nabla^2w-n\text{cos}(ny),\\&\quad (t,x,y)\in[0,\infty)\times[0,2\pi]\times[0,2\pi]\\
w(0,x,y) &= w^0(x,y),\quad (x,y)\in[0,2\pi]\times[0,2\pi] 
\end{aligned}
\end{equation}
We use a wave number of the forcing function $n=4$ and a Reynolds number of $Re=40$.

\textbf{Dataset Generation. } The training dataset consists of 160 trajectories, each simulated for 500 seconds with snapshots saved every 1 second. Each snapshot has a spatial resolution of 64 × 64. The validation set contains 40 trajectories with the same discretization.

\textbf{Model Architecture. } The DeepONet backbone consists of a branch network with four convolutional layers (output dimensions 64, 128, 256, 512) and two fully connected layers (1024 neurons each). The trunk network consists of five fully connected layers (1024 neurons each).

\section{Future Work}
Future work could explore more expressive non-quadratic energy functionals, potentially even parameterized by another neural network, and extend the projection framework to enforce other nonlinear physical constraints. Investigating the computational scaling of this approach to even larger-scale and realistic systems also remains a promising direction for building reliable and physically consistent dynamics emulators.



\end{document}